\documentclass[aps,prl,twocolumn,showpacs,superscriptaddress]{revtex4}
\usepackage{amssymb}

\usepackage{color}
\usepackage{amsmath}
\usepackage{graphicx}
\usepackage{bm}

\begin{document}

\title{Quantum Electrodynamical Photon Splitting in Magnetized Nonlinear Pair Plasmas}

\author{G. Brodin} 
\affiliation{Department of Physics, Ume\aa\ University, SE-901 87 Ume{\aa},
Sweden} 

\author{M. Marklund}
\affiliation{Department of Physics, Ume\aa\ University, SE-901 87 Ume{\aa},
Sweden} 

\author{B. Eliasson} 
\affiliation{Department of Physics, Ume\aa\ University, SE-901 87 Ume{\aa},
Sweden} 
\affiliation{Institut f\"ur Theoretische Physik IV, Ruhr-Universit\"at Bochum, 
D-44780 Bochum, Germany}

\author{P. K. Shukla}
\affiliation{Department of Physics, Ume\aa\ University, SE-901 87 Ume{\aa},
Sweden} 
\affiliation{Institut f\"ur Theoretische Physik IV, Ruhr-Universit\"at Bochum, 
D-44780 Bochum, Germany}

\begin{abstract}
We present for the first time the nonlinear dynamics of quantum
electrodynamic (QED) photon splitting in a strongly magnetized
electron-positron (pair) plasma. By using a QED corrected Maxwell equation,
we derive a set of equations that exhibit nonlinear couplings between
electromagnetic (EM) waves due to nonlinear plasma currents and QED
polarization and magnetization effects. Numerical analyses of our coupled
nonlinear EM wave equations reveal the possibility of a more efficient decay
channel, as well as new features of energy exchange among the three EM modes
that are nonlinearly interacting in magnetized pair plasmas. Possible
applications of our investigation to astrophysical settings, such as
magnetars, are pointed out.
\end{abstract}

\received{3 May, 2006} 
\revised{19 August, 2006, and 19 December, 2006}

\maketitle

Recently, there has been a great deal of interest \cite
{marklund-shukla} in the investigation of effects
associated with radiation pressure and quantum vacuum fluctuations in
nonlinear media. Such studies are of importance in astrophysical
environments, 
where copious
amounts of electron-positron pairs exist due to numerous physical processes
\cite{marklund-shukla}. Elastic photon-photon
scattering is traditionally described within quantum
electrodynamic (QED) \cite{marklund-shukla,book}.
However,
observable effects of elastic photon--photon scattering among real photons
have so far not been reported in the laboratory \cite{marklund-shukla,Moulin1999-BMS2001}. 
For astrophysical systems 
\cite{Harding,Marklund-Brodin-Stenflo} the situation is different, since the
large magnetic field strength in pulsar and magnetar \cite{magnetar}
environments changes the diamagnetic properties of vacuum significantly \cite{magnetic}, 
and leads to phenomena such as frequency down-shifting \cite
{Harding}. The latter is a result of photon splitting \cite
{Bialynicki-Birula1970,Adler}, and the process may be responsible for the
radio silence of magnetars \cite{Harding,Baring-Harding}. 
Moreover, the propagation of electromagnetic waves 
in a relativistically dense electron gas \cite{electron} and in a relativistic electron--positron gas 
\cite{pair} have been discussed, leading to the important effect of gamma photon capture and pair plasma
suppression around pulsars 
\cite{shabad}. 

In this Letter, we present the nonlinear photon splitting
of electromagnetic (EM) waves propagating perpendicularly to a strong external
magnetic field $\mathbf{B}_{0}$ in a pair plasma. Due to the QED effect \cite
{marklund-shukla}, a photon in vacuum can decay into a backscattered and a
forward scattered photon, where the latter two photons have polarizations
perpendicular to that of the original photon \cite
{Bialynicki-Birula1970,Adler}. Noting that significant pair-production \cite
{Beskin-book} occurs in astrophysical settings (viz.\ in pulsar and magnetar
environments), we demonstrate here a novel possibility of a nonlinear decay
interaction, due to a competition between 
QED and plasma nonlinearities. We note that most of previous
investigations \cite{lai-ho,
brodin-etal}, including
both QED and plasma effects, have been limited to linear EM wave
propagation. Here we derive three dynamical equations with nonlinear couplings between photons with different polarizations.  
From these coupled mode equations, the QED
cross-section for photon-splitting \cite{Bialynicki-Birula1970} can be
deduced in the limit of zero plasma density. 
We discuss applications of our results 
to magnetar atmospheres.

Photon-photon scattering 
can be described by the Heisenberg--Euler
Lagrangian \cite{Heisenberg-Euler,Schwinger} $L = \epsilon_0(E^2 - c^2B^2)/2 + \kappa
\epsilon_0^2[ (E^2 - c^2B^2)^2 + 7(c\mathbf{E}\cdot\mathbf{B})^2]$. Here $\kappa =(\alpha /90\pi )(1/\epsilon _{0}E_{\text{crit%
}}^{2})$, $\alpha = e^2/4\pi\epsilon_0\hbar c$ is the fine-structure constant, $E_{\text{crit}%
}=m^{2}c^{3}/e\hbar \sim 10^{18}\,\mathrm{V/m}$ is the critical
field \cite{marklund-shukla}, $\hbar $ is the Planck constant, $m$ is the electron mass, $\epsilon_0$ is the vacuum permittivity,
and $c$ is the vacuum speed of light. The last two terms in the
Heisenberg--Euler Lagrangian represent the effects of the vacuum
polarization and magnetization. The QED corrected Maxwell equations can then
be written in their classical form using $\mathbf{D}=\epsilon _{0}\mathbf{E} + 
\mathbf{P}\ $ and $\mathbf{H} = c^2\epsilon_0\mathbf{B} - \mathbf{M}$, where 
\cite{marklund-shukla} 
$\mathbf{P}=2\epsilon_0^2\kappa[ 2(E^{2}-c^{2}B^{2})\mathbf{E}+7c^{2}(%
\mathbf{E}\cdot \mathbf{B})\mathbf{B}]$ and 
$\mathbf{M}= 2c^2\epsilon_0^2\kappa[ -2(E^{2}-c^{2}B^{2})\mathbf{B}+7(%
\mathbf{E}\cdot \mathbf{B})\mathbf{E}]$, 
which are valid for $|\mathbf{E}|, c|\mathbf{B}|\ll E_{\text{crit}}$ and $%
\omega \ll \omega_{e} = mc^{2}/\hbar \approx 8\times
10^{20}\, \mathrm{rad/s}$. 

Next, we study wave propagation perpendicular to an external
magnetic field $\mathbf{B}_{0}=B_{0}\widehat{\mathbf{z}}$ in an
electron-positron plasma, letting all variables depend on $(x,t)$. 
Assuming that the charge density is negligible, the
wave equation for the electric field $\mathbf{E}$ then reads 
\begin{equation}
\left( \partial_t^2 - c^2\partial_x^2\right) \mathbf{E} = -
\epsilon_0^{-1}\left[ \partial_t\mathbf{j} - c^2\widehat{\mathbf{x}}\partial_x^2P_x \right],
\label{Wave-eqs}
\end{equation}
where $\mathbf{j} = \partial_t\mathbf{P}+\nabla \times \mathbf{M} +
\sum_{e,p}qn\mathbf{v}$, $|\mathbf{E}|/c,|\mathbf{B}|\ll B_{0}$, $\mathbf{v}$ denotes the average (fluid) velocity,  and the sum
is over the electron and positron contributions. The latter are determined
from the relativistic equation of motion
\begin{equation}
\left( \partial_t + \mathbf{v\cdot \nabla }\right) (\gamma \mathbf{v)} =
(q/m)\left( \mathbf{E+v}\times \mathbf{B}\right) .  \label{Motion-eqs}
\end{equation}
We assume that the EM wave frequency $\omega $ and the electron (positron)
plasma frequency $\omega _{pe(p)}$ are much smaller than the magnitude of
the electron (or positron) gyrofrequency $\left| \omega _{ce(p)}\right| = \omega_c = eB_0/m$,
relevant for pulsar and magnetar atmospheres \footnote{
  The cold collisionless momentum equation is valid for
  $\nu_{\rm ep} \ll \omega_p$, $v_{\rm th} \ll c$, conditions 
  fulfilled for a broad range of densities and temperatures. 
  Here $\nu_{\rm ep}$ is the collision frequency and $v_{\rm th}$ is the thermal velocity.
  }. 
This ordering will make charge density oscillations negligible, as the longitudinal 
motion for both the electrons and positrons will be given by the $\mathbf{E}\times \mathbf{B}$-drift
to leading order. Next, we
linearize and Fourier decompose Eqs.\ (\ref{Wave-eqs}) and (\ref{Motion-eqs}) 
to obtain the dispersion relations for the EM waves
propagating perpendicular to $\widehat{\mathbf{z}}$ \footnote{
  The dispersion relation is much simpler for perpendicular 
  propagation, but the main features remain for 
  arbitrary propagation directions (cf.\ (12) and (13) in Ref.\ \cite{brodin-etal}):
  the combined QED/plasma 
  effect is important for one of the polarizations, while only QED affects the other.
} 
\begin{subequations}
\begin{eqnarray}
  && \omega ^{2} \approx k^{2}c^{2}\left( 1-8\xi \right) ,  \label{MS-mode} \\
  && \omega ^{2} \approx k^{2}c^{2}(1-14\xi )+\omega _{p}^{2},  \label{ord-mode}
\end{eqnarray}
\end{subequations}
where $\omega _{p}=(\omega _{pe}^{2}+\omega _{pp}^{2})^{1/2}$ is the plasma
frequency of the pair plasma. Here, $\xi =\kappa \epsilon
_{0}c^{2}B_{0}^{2} \ll 1$, and we assume that $\omega _{p}^{2}\ll \omega
^{2}$ \footnote{
  While accurate in the regime $\omega_p \ll \xi\omega_{\perp}$, 
  the linear 
  QED contribution  
  in principle needs to be modified for the lf-mode when $\omega _{p} \gtrsim\xi \omega _{\bot }$.
  However, the plasma contribution dominates the linear regime 
  in this case, as $\xi \ll 1$, and Eqs.\ (3) and (\ref{Coupling}) are thus sufficient also
  when $\omega \rightarrow \omega_p $.
}. We have omitted the contribution proportional to $\omega
^{2}\omega _{p}^{2}/\omega _{c}^{2}$ in (\ref{MS-mode}), 
which is smaller than the plasma contribution
proportional to $\omega _{p}^{2}$ in (\ref{ord-mode}). We note that the EM
mode described by (\ref{MS-mode}) has the electric field perpendicular to $%
\mathbf{B}_{0}$ in the $\widehat{\mathbf{y}}$-direction (approximately), whereas 
the EM mode described by (\ref{ord-mode}) has the
electric field parallel to $\mathbf{B}_{0}$. Henceforth, the two different
polarizations will be denoted by the subscripts $\bot $ and $\Vert $,
respectively. In \cite{brodin-etal} and \cite{Marklund2006}, the
linear effects from the combined QED and plasma effects are discussed in detail.

Next, we represent the EM waves as $\widetilde{E}(x,t)\exp (ikx-i\omega t)$
+ complex conjugate, and the slowly varying amplitudes are denoted by tilde.
Our aim is to investigate photon splitting \cite
{Bialynicki-Birula1970,Adler}, a parametric process where one
photon with perpendicular polarization decays into two photons with parallel
polarizations. Denoting the latter waves with indices 1 and 2, the
energy and momentum conservation relations (matching conditions) are $\omega
_{\bot }=\omega _{1\Vert }+\omega _{2\Vert }$ and $k_{\bot }=k_{1\Vert
}+k_{2\Vert }$. We point out that in addition to this process, QED allows a
decay of the type $\omega _{\bot }=\omega _{1\bot }+\omega _{2\Vert }$ \cite
{book}. However, the scattering amplitude of this process is suppressed by a
factor of the order $\alpha \xi $ \cite{book}. The presence of a plasma may in principle
change this ordering and also add new decay channels, but in the strongly
magnetized high-frequency regime considered here, $\omega ,\omega _{c}\gg
\omega _{p}$, we note that this is not the case. For $\omega _{p}=0$, the
simultaneous fulfilment of the matching conditions and the dispersions
relations (\ref{MS-mode}) and (\ref{ord-mode}) requires that one of the EM
waves is backscattered, i.e. either $k_{1\Vert }<0$ or $k_{2\Vert }<0$.
Furthermore, $\xi \ll 1$ means that the backscattered EM wave has a much
smaller frequency than $\omega _{\bot }$. On the other hand, for $\omega
_{p}\neq 0$, the matching conditions also allow for both the decay products
to be scattered in the forward direction. 
Next, we divide all quantities into unperturbed and perturbed quantities, i.e. $\mathbf{B}%
=\mathbf{B}_{0}+\mathbf{B}_{1}$ where the perturbed part fulfills $\left| 
\mathbf{B}_{1}\right| \ll \left| \mathbf{B}_{0}\right| $, and similarly for
the polarization, magnetization and density. We then include the resonant
second order nonlinear terms from $\mathbf{P}$ and $\mathbf{M}$ in Maxwell's
Eqs. (noting that $\mathbf{P=P}_{0}+\mathbf{P}_{1}+\mathbf{P}_{2}$, where $%
\mathbf{P}_{2}$ is second order in the perturbed EM-field, etc.) together
with the second order terms in (\ref{Motion-eqs}) (i.e. the Lorentz force
and the convective derivative), and the nonlinear part of the current
density in the right-hand side of (\ref{Wave-eqs}).
From the continuity equation $\partial_t\rho_1 = -\rho_0\partial_xv_x$ we
solve for the density. After
straightforward algebra, where the linear expressions are substituted into
the nonlinear terms, we obtain our coupled mode equations 
\begin{subequations}
\begin{eqnarray}
&&\!\!\!\!\!\!\!\!\!\!\partial _{t}\widetilde{E}_{\bot }+v_{g\bot }\partial
_{x}\widetilde{E}_{\bot }=\omega _{\bot }C\widetilde{E}_{1\Vert }\widetilde{E%
}_{2\Vert }/E_{\mathrm{crit}},  \label{Coupling1} \\
&&\!\!\!\!\!\!\!\!\!\!\partial _{t}\widetilde{E}_{1\Vert }+v_{g1}\partial
_{x}\widetilde{E}_{1\Vert }=\omega _{1\Vert }C\widetilde{E}_{2\Vert }^{\ast }%
\widetilde{E}_{\bot }/E_{\mathrm{crit}},  \label{Coupling2} \\
&&\!\!\!\!\!\!\!\!\!\!\partial _{t}\widetilde{E}_{2\Vert }+v_{g2}\partial
_{x}\widetilde{E}_{2\Vert }=\omega _{2\Vert }C\widetilde{E}_{1\Vert }^{\ast }%
\widetilde{E}_{\bot }/E_{\mathrm{crit}},  \label{Coupling3}
\end{eqnarray}
\label{Coupling}
\end{subequations}
where $v_{gj}=\partial \omega _{j}/\partial k_{j}$ is the group speed ($j$
equals $\bot ,\,1\Vert $ and $2\Vert $), and the
asterisk denotes complex conjugate. The coupling strength is $C=C_{\mathrm{%
pl}}+C_{\mathrm{QED}}$, where
$C_{\mathrm{pl}}=\mathrm{i}\left( {\alpha }/{90\pi \xi }\right) ^{1/2}%
({k_{\bot }c}/{\omega _{\bot }})({\omega _{p}^{2}}/{\omega _{1\Vert
}\omega _{2\Vert }})$ is due to the plasma
nonlinearities \footnote{
  The plasma nonlinearities are related to a nonlinear conductivity
  $\bm{\mathsf{\sigma}}_{\rm nl}$, defined by $j_{\mathrm{nl}\,\perp} 
    = \sigma_{\mathrm{nl}\,\perp 1 2}E_{1\|}E_{2\|}$ etc., i.e.\
  $\sigma_{\mathrm{nl}\,\perp 1 2} = \omega_{\perp}\epsilon_0C_{\mathrm{pl}}/E_{\mathrm{crit}}$.
} 
and  
$C_{\mathrm{QED}}=2\mathrm{i}\left( {\alpha \xi }/{90\pi }\right) ^{1/2}%
\left[ 10 ({k_{\bot }c}/{\omega _{\bot }})+7\left( {k_{1\Vert }c}/{%
\omega _{1\Vert }}+{k_{2\Vert }c}/{\omega _{2\Vert }}\right) \right]$ is due to QED nonlinear 
interactions 
\footnote{
  We note that $\lim_{\hbar\rightarrow 0}(C_{\mathrm{pl}}/E_{\mathrm{cr}})$ is finite [cf.\ (\ref{Coupling})].
}.
During certain conditions and for sufficiently long times, the multi-scale expansion behind Eqs.\ 
(\ref{Coupling}) can break down due to the growth of higher order terms. However, we
will assume that, due to convective stabilization (see below), such effects are negligible 
and the evolution is well described by (\ref{Coupling}). 
 

Let us now discuss the relative importance of the QED and plasma effects in
various regimes.

\textit{Case 1:}\ In the regime $\omega _{p}/\omega
_{\bot }\ll \xi $, the matching conditions, the dispersion relations, and $%
\xi \ll 1$ give $k_{2\Vert } 
\approx -3\xi k_{\bot }/2$ and $\omega _{2\Vert }\approx 3\xi \omega _{\bot
}/2$, where the last approximation is valid to first order in $\xi $, and we
have chosen $\omega_{2\Vert }$ as the low-frequency backscattered wave with $%
k_{2\Vert }<0$. We note that the condition $\omega _{p}/\omega _{\bot }\ll
\xi $ ensures that the QED effect dominates over the plasma effects.

\textit{Case 2:}\ Increasing the plasma density starts to affect the linear
properties of the low-frequency EM mode first. However, we note that the
expression $\omega _{2\Vert }=3\xi \omega _{\bot }/2$ even holds when the
plasma effect dominates over the QED effect in the dispersion relations for
the low-frequency EM mode. Increasing the plasma frequency to the regime $%
\omega _{p}/\omega _{\bot }\sim $ $\xi $, the nonlinear coefficients $C_{%
\mathrm{QED}}$ and $C_{\mathrm{pl}}$ become comparable and the matching
conditions further imply that $\omega _{2\Vert }$ $\sim \omega _{p}\sim $ $%
\xi \omega _{\bot }$ [see the note after Eqs.\ (3)]. Furthermore, we note that for $%
\omega _{p}\gtrsim 3\xi \omega _{\bot }$ we would have forward scattering
instead of backscattering of the low-frequency EM mode.

\textit{Case 3:}\ For larger plasma frequency ($\omega _{p}/\omega
_{\bot}\gg \xi $) the QED-contribution to the frequency $\omega _{\bot }$
may still dominate over the plasma contribution, but $|C_{\mathrm{QED}}|\ll
|C_{\mathrm{pl}}|$, and for the nonlinear wave interaction we can thus omit
the QED effect in this regime.

We note that the nonlinear system (\ref{Coupling})
has the conserved energy integral $\mathcal{E}=\int (|E_{\bot
}|^{2}+|E_{1\Vert }|^{2} +|E_{2\Vert }|^{2})\,dx$ when $\omega _{\bot
}=\omega _{1\Vert }+\omega _{2\Vert }$, and that the two other linearly
independent constants of motion are $\mathcal{N}_{1}=\int (|E_{\bot
}|^{2}/\omega _{\bot }-|E_{1\Vert }|^{2}/\omega _{1\Vert })\,dx$ and $%
\mathcal{N}_{2}=\int (|E_{\bot }|^{2}/\omega _{\bot }-|E_{2\Vert
}|^{2}/\omega _{2\Vert })\,dx$, corresponding to the Manley-Rowe relations.
The constants of motion are used as a check of the numerical calculations
presented below. We
first make a linear stability analysis in the presence of a pump EM wave.
Thus, we consider the decay of a homogeneous intense wave $E_{\bot }=E_{\bot
0} $, where $|E_{\bot 0}|\gg |E_{1\Vert }|,\,|E_{2\Vert }|$, into daughter
EM waves $E_{1\Vert }=\widehat{E}_{1\Vert }\exp (iKx-i\Omega t)$ and $%
E_{2\Vert }=\widehat{E}_{2\Vert }\exp (-iKx+i\Omega t)$. From (\ref
{Coupling}), we obtain the nonlinear dispersion relation 
$(\Omega -v_{g2\Vert }K)(\Omega -v_{g1\Vert }K) =-\omega _{1\Vert }\omega
_{2\Vert }|C|^{2}|E_{\perp 0}|^{2}/E_{\mathrm{crit}}^{2}$. From the latter,
the growth rate of the daughter EM waves is obtained as $\Gamma =[\omega
_{1\Vert }\omega _{2\Vert }|C|^{2} {|E_{\perp 0}|^{2}}/{E_{\mathrm{crit}}^{2}%
} - {(v_{g1\Vert }+v_{g2\Vert })^{2}K^{2}}/4]^{1/2}$ for wavenumbers $%
K<2(\omega _{1\Vert }\omega _{2\Vert })^{1/2}|C E_{\perp 0}|/|v_{g2\Vert }
+v_{g1\Vert }|E_{\mathrm{crit}}$.


\begin{figure}[tbp]
\centering
\includegraphics[width=0.9\columnwidth]{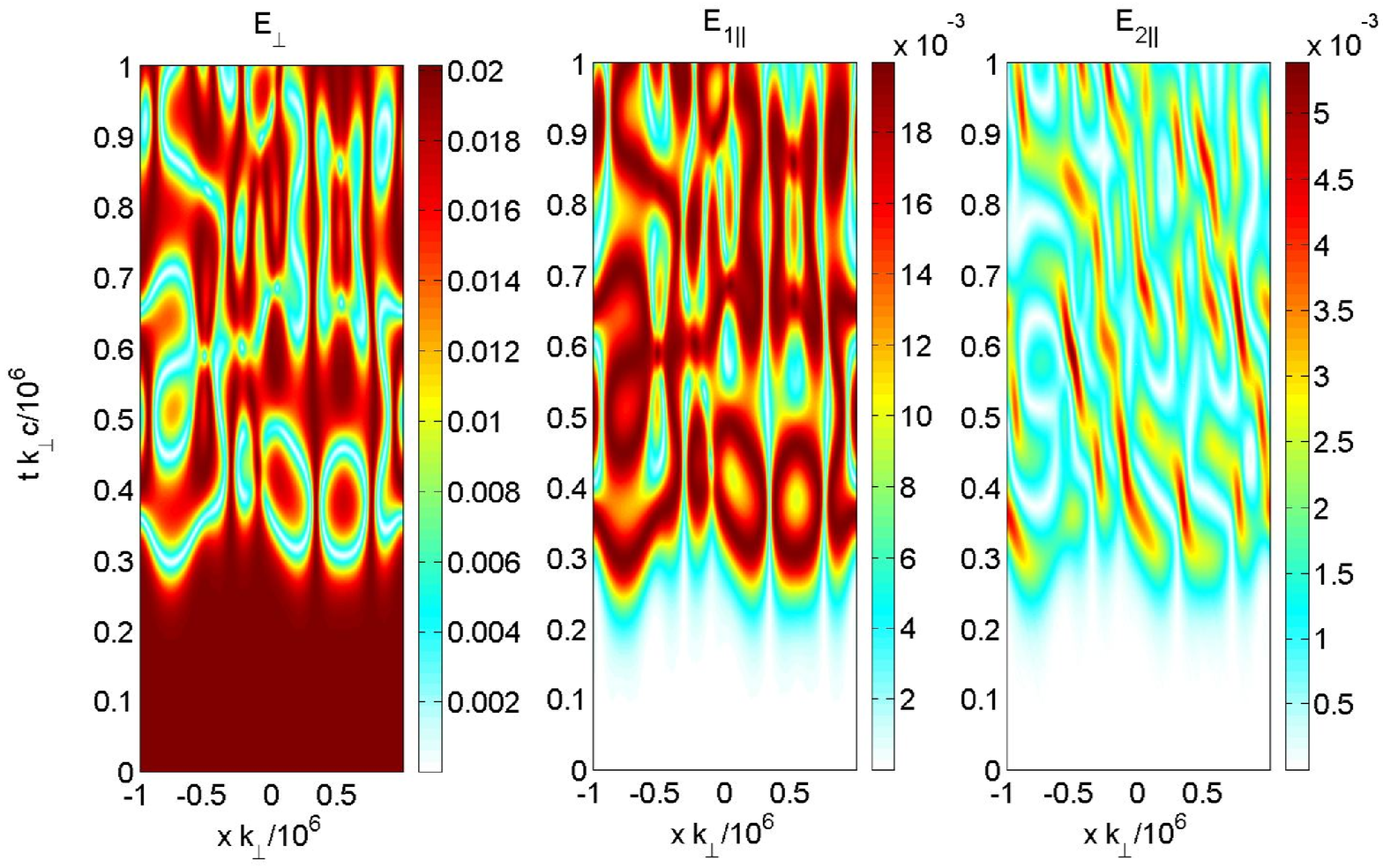}
\caption{The decay of the pump mode $E_{\perp}=\widetilde{E}_\perp/E_{%
\mathbf{crit}}$ into a forward scattered mode $E_{1\Vert}=\widetilde{E}%
_\perp/E_{\mathbf{crit}}$ and a back-scattered mode $E_{2\Vert}=\widetilde{E}%
_\perp/E_{\mathbf{crit}}$. Initially, the pump is set to $E_\perp=0.02$,
while $E_1$ and $E_2$ is set to a low-level random noise. After the
initial exponential decay, the energy
is transferred between the pump and the two EM sidebands in a semi-periodic
and chaotic manner. We used $\protect\omega_p=0$ and $\protect\xi=0.01$,
yielding $k_{1\Vert}=1.017k_\bot$, $k_{2\Vert}=-0.017k_\bot$, $\protect\omega%
_\bot=0.959k_\perp c$, $\protect\omega_{1\Vert}=0.943 k_\perp c$, and $\protect%
\omega_{2\Vert}=0.016k_\perp c$. }
\end{figure}

\begin{figure}[tbp]
\centering
\includegraphics[width=0.9\columnwidth]{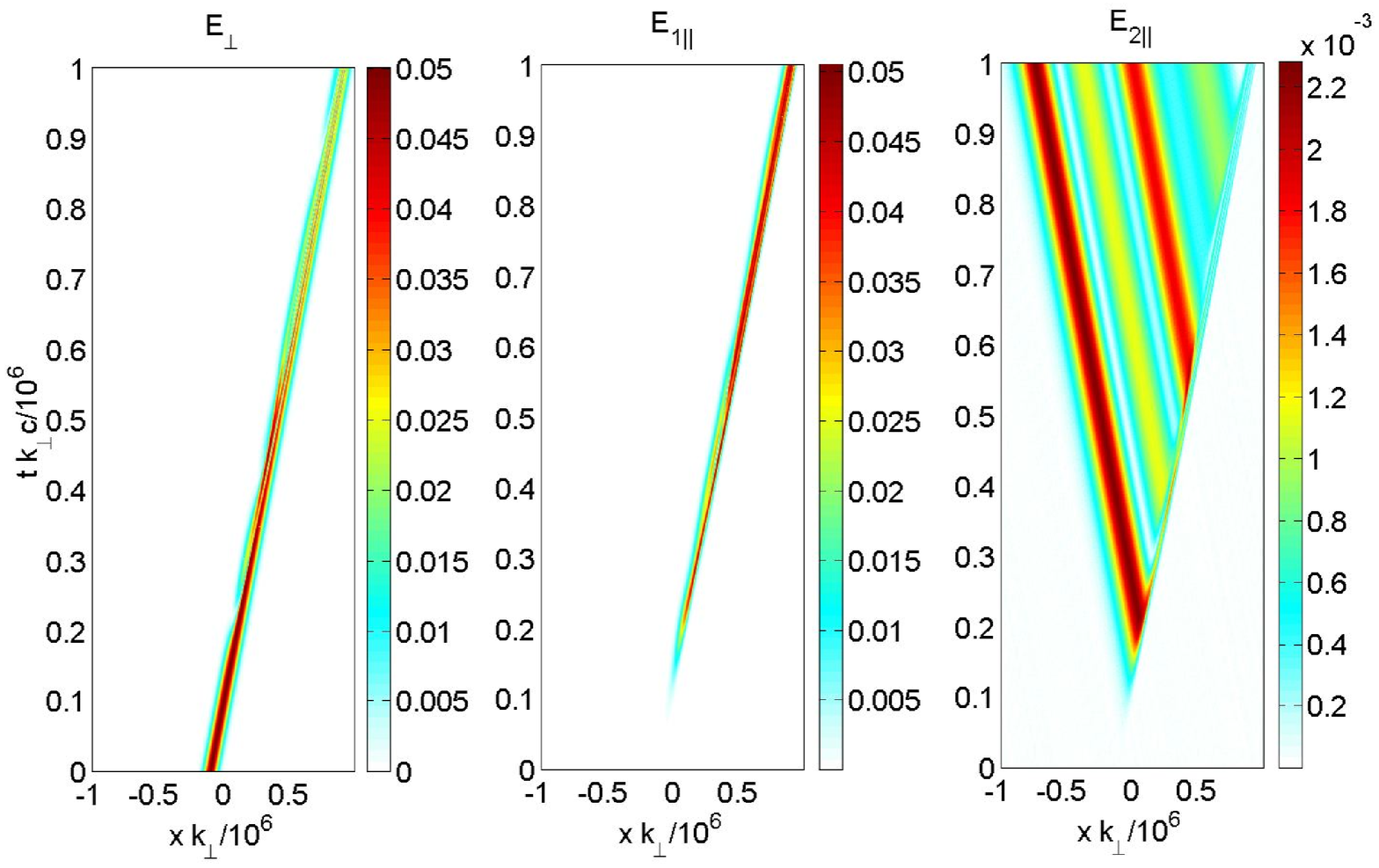}
\caption{The decay of the pump mode $E_{\perp}=\widetilde{E}_\perp/E_{%
\mathbf{crit}}$. The pump was initially set to a localized pulse $%
E_\perp=0.05\exp[-(xk_\perp-10^5)^2/2.5\times10^9]$. The pump decays into a
forward scattered sideband $E_{1\Vert}$ and a back-scattered sideband $%
E_{2\Vert}$. We used $\protect\omega_p=0$ and $\protect\xi=0.01$, yielding $%
k_{1\Vert}=1.017k_\bot$, $k_{2\Vert}=-0.017k_\bot$, $\protect\omega%
_\bot=0.959k_\perp c$, $\protect\omega_{1\Vert}=0.943 k_\perp c$, and $\protect%
\omega_{2\Vert}=0.016k_\perp c$. }
\end{figure}

\begin{figure}[tbp]
\centering
\includegraphics[width=0.9\columnwidth]{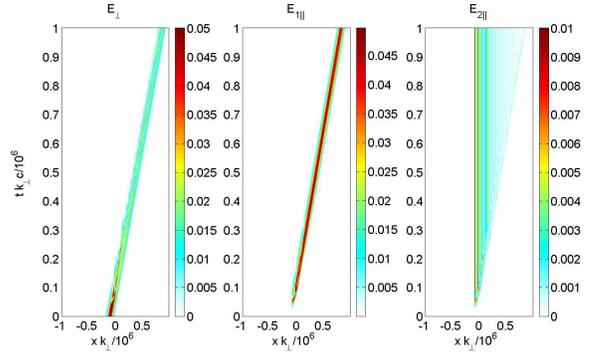}
\caption{The decay of the pump mode $E_{\perp}=\widetilde{E}_\perp/E_{%
\mathbf{crit}}$ in a strongly magnetized plasma. The pump was initially set
to a localized pulse $E_\perp=0.05\exp[-(xk_\perp-10^5)^2/2.5\times10^9]$.
The pump decays into a forward scattered sideband $E_{1\Vert}$ and a
sideband $E_{2\Vert}$, which has almost zero group speed. We used $\protect%
\xi=0.01$ and $\protect\omega_p=3\protect\xi\protect\omega_\perp=0.03\protect%
\omega_\perp$, yielding $k_{1\Vert}=k_\bot$, $k_{2\Vert}\approx 0$, $\protect%
\omega_\bot=0.959 k_\perp c$, $\protect\omega_{1\Vert}=0.929 k_\perp c$, and $%
\protect\omega_{2\Vert}=\protect\omega_p\approx 0.030 k_\perp c$. }
\end{figure}


In order to study the dynamics of an intense electromagnetic beam in a
strong magnetic field, we numerically solve the coupled Eqs.\ (\ref
{Coupling}), and display the results in Figs.\ 1--3. 
We use $1000$ grid points to resolve the numerical domain $-10^6 \leq x k_\perp \leq 10^6$
with periodic boundary conditions, and $20\,000$ steps to advance the solution
in time. A pseudo-spectral method is used to approximate the spatial derivatives
and a $4^{\rm th}$-order Runge-Kutta method for the time stepping.
In Figs.\ 1 and 2, the plasma is absent, so that the only EM
wave couplings are due to the QED effect. We used $\xi =0.01$, yielding $k_{1\Vert }=1.017k_{\bot }$, $k_{2\Vert
}=-0.017k_{\bot }$, $\omega _{\bot }=0.959k_{\perp }c$, $\omega _{1\Vert
}=0.943k_{\perp }c$, and $\omega _{2\Vert }=0.016k_{\perp }c$. In Fig.\ 1, we
present the evolution of an initially homogeneous beam of
amplitude $\widetilde{E}_{\perp }=0.02\,E_{\mathrm{crit}}$. Initially, the two daughter waves grow exponentially,
followed by a nonlinear oscillatory phase. Figure 2 exhibits the nonlinear
dynamics of a localized wave packet. We observe the decay
of the EM pulse into a forward scattered wave $E_{1\Vert }$ and a
backscattered wave $E_{2\Vert }$. In Fig.\ 3, we show the dynamics of a localized EM pulse when
the plasma effect is important. We consider the
particular case $\omega _{p}=3\xi \omega _{\perp }$, so that the
low-frequency daughter EM wave $E_{2\Vert }$ has approximately zero group
speed and a frequency that equals the plasma frequency. We thus used $%
\xi =0.01$ and $\omega _{p}=3\xi \omega _{\perp }=0.03\omega _{\perp }$,
yielding $k_{1\Vert }=k_{\bot }$, $k_{2\Vert }\approx 0$, $\omega _{\bot
}=0.959k_{\perp }c$, $\omega _{1\Vert }=0.929k_{\perp }c$, and $\omega
_{2\Vert }=\omega _{p}\approx 0.030k_{\perp }c$. The energy
of the pump $E_{\perp }$ is transferred to a forward scattered 
wave $E_{1\Vert }$ and zero-group speed waves $%
E_{2\Vert }$. 

The present investigation is of relevance for EM wave propagation
in the vicinity of pulsars and magnetars. For
example, the radio silence of magnetars is assumed to be connected with the
photon-splitting in the strong magnetar fields ($10^{9}-10^{11}\,\mathrm{T}$) 
\cite{Baring-Harding,Harding}. Photon splitting can suppress the
creation of electron--positron pairs \cite{Baring-Harding}, but we still expect
the presence of an electron--positron plasma \cite{Beskin-book} in such
environments. The
Goldreich--Julian density is given by \cite{Goldreich-Julian} $n_{GJ}=7\times 10^{15}({0.1}%
/{\tau })({B_{p}}/{10^{8}})\,\mathrm{m}^{-3}$, where $\tau $ is the pulsar
period time (in seconds) and $B_{p}$ is the surface pulsar magnetic field
(in Tesla). The pair plasma density is expected to satisfy $%
n_{e}=n_{p}=Mn_{GJ}$, where a moderate estimate of the multiplicity 
gives $M=10$ \cite{Asseo}. Choosing this value and
letting $\tau =1\,\mathrm{s}$, we note that for the weak magnetar field
strength $B_{p}=10^{9}\,\mathrm{T}$, the characteristic pump frequency $%
\omega _{\bot \mathrm{char}}\sim \omega _{p}/\xi $, where the QED and plasma
effects are of equal importance when the photon splitting process is of the
order of $\omega _{\bot \mathrm{char}}\sim 4\times 10^{15}\,\mathrm{rad/s}$,
i.e.\ in the optical range. For $\omega _{\bot }\ll \omega _{\bot \mathrm{char%
}}$, the plasma nonlinearities dominate, whereas the QED effect dominates in
the opposite regime. The evolution of the coupled system of Eqs.\ (\ref
{Coupling}) is, to a large extent, controlled by the
pulse length of the pump mode. 
For long
pulse-lengths with $L\gg (\omega _{\bot }CE_{\bot }/E_{\mathrm{crit}})^{-1}$
(Fig.\ 1), the system shows a ``predator-prey'' type of behavior where
the energy oscillates chaotically between the different modes. For a
moderate pulse-length with $L\sim (\omega _{\bot }CE_{\bot }/E_{\mathrm{crit}%
})^{-1}$, the excited EM wave energy propagates out of the interaction
region, and one encounters a somewhat more ordered behavior and an effective 
damping of the pump mode. The EM wave
energy is then mainly converted to the parallel polarized forward scattered
EM mode (Fig.\ 2). For a short pulse-length with $L\ll (\omega _{\bot
}CE_{\bot }/E_{\mathrm{crit}})^{-1}$, thermal fluctuations do not grow
due to convective stabilization (the growth rate $\Gamma $ vanishes for
large wavenumbers), and effectively the nonlinear interaction vanishes.
As the plasma density increases, plasma effects become important,  
as depicted in Fig.\ 3. For $\omega
_{p}\sim \xi \omega _{\bot \mathrm{char}}$, there are two simultaneous effects of the
plasma. First, in this regime
the QED and plasma contributions to the coupling strength are comparable,
i.e.\ $C_{\mathrm{pl}}\sim C_{\mathrm{QED}}$, which increases the total
coupling strength $C=C_{\mathrm{pl}}+C_{\mathrm{QED}}$, since the phases of $%
C_{\mathrm{pl}}$ and $C_{\mathrm{QED}}$ coincide. Second, when the group
velocity of the backscattered EM wave is slowed down, the effectiveness of
convective stabilization is diminished, increasing the interaction strength
and speeding up the conversion of the EM wave energy. Similarly to the case
without the plasma (Fig.\ 2), the EM wave energy mainly ends up in the
parallel polarized forward scattered EM mode, but the characteristic
splitting timescale is considerably faster with the plasma present. 
We stress that the simulations presented in this Letter contain results
that are easily generalizable to other parameter ranges, following the discussion presented above. The
combined effect of plasma and QED effects should have consequences for the
emission spectra from magnetars and pulsars. While the QED effects alone
shifts the spectrum towards linear polarization, we emphasize that the
effect is much more pronounced when plasma effects are present, which holds
for radiation with frequencies of the order $\omega _{\mathrm{char}}\sim
\omega _{p}/\xi $, i.e.\ in the optical range for
magnetars, while in the infrared to microwave range for pulsars. Thus, we
suggest that evidence for a combined plasma-QED photon splitting process
should be sought for in the polarization signature of magnetar and pulsar
emission.

\end{document}